# Context-Aware Generative AI for Automated Telecom Test Script Generation

**Gautam Prasad, Chandramohan T. N., Joy Bose**

*Ericsson R&D Bangalore*

{gautam.prasad, chandramohan.tn, joy.bose}@ericsson.com

**Abstract—**Automated test generation for telecom software systems and networks has advanced significantly with the adoption of machine learning and rule-based approaches. However, most existing solutions generate static test suites against a snapshot of the system; as code, configurations, topologies, and key performance indicators (KPIs) evolve, these tests quickly become outdated or misaligned with the live system. There is currently no widely adopted solution that continuously detects fine-grained changes and selectively adapts only the affected tests without regenerating entire test suites. This paper presents a context-aware generative AI framework for automated telecom test script generation that treats testing as a continuously adapting process driven by the current state of the system rather than a static artifact. The central contribution is delta-conditioned test generation over a live knowledge graph: our approach employs a continuously updated knowledge graph (KG) as a single source of truth, a delta engine for fine-grained change detection, and a KG-guided generative AI agent, operating via the Model Context Protocol (MCP), to create, update, or retire test cases automatically. We further integrate Retrieval-Augmented Generation (RAG) to enrich reasoning with telecom-domain knowledge and historical artifacts. We demonstrate applicability across software-system and telecom-network use cases, including a Python-based KPI monitoring application managed in GitLab, and show how the framework reduces manual effort, improves test relevance, and accelerates test cycles.

## 1 Introduction

Telecom networks and the software systems that manage them are in a state of constant evolution. New features are introduced, configurations are updated, KPIs drift, and network topologies change due to scaling or optimization activities. 5G and beyond networks have introduced unprecedented complexity: cloud-native Network Functions (NFs) deployed on Kubernetes, O-RAN disaggregation introducing multi-vendor interoperability requirements, and network slicing creating dynamic, per-slice SLA obligations. Each of these dimensions demands that test suites remain precisely aligned with the current system state, which is a requirement that static, snapshot-based testing approaches fundamentally cannot satisfy.

While automated test generation tools exist for both software and network domains, they typically rely on static rules, templates, or one-time learning from historical data. As a result, generated test suites often become misaligned with the live system unless they are manually regenerated or curated, leading to wasted testing effort, missed defects, and slower development and deployment cycles. Industry surveys consistently report that test maintenance consumes between 30% and 50% of overall software testing effort [1], a figure that is likely higher in telecom environments given the pace of network evolution.

The emergence of large language models (LLMs) and generative AI has opened new possibilities for automated test synthesis [2, 3], while knowledge graphs have demonstrated their value as structured, queryable repositories of domain knowledge in network management and operations [4, 5]. Yet these capabilities have not been unified into a framework that (a) continuously tracks system state, (b) detects meaningful changes at fine granularity, and (c) generates or adapts tests selectively in response to those changes.

Currently, there is no widely adopted solution that simultaneously (1) maintains an incrementally updated representation of both network and software context, (2) detects fine-grained changes in this representation, and (3) dynamically generates or adapts test cases in response to those changes. Existing automated testing tools typically generate static test suites and do not exploit a unified, knowledge-graph-based view of the system under test.

In this paper, we propose a context-aware generative AI framework for automated telecom test script generation that addresses these limitations. The key idea is to treat testing as a continuously adapting process driven by the current state of the system under test rather than a one-off activity. We ground this approach in the observation that a knowledge graph, when kept current through automated ingestion, naturally captures the structural and semantic relationships that determine test relevance, making it the ideal substrate for driving generative test synthesis.

Our main contributions are:

1. **Knowledge-graph-based representation of telecom and software context** that unifies topology, KPIs, configurations, test artifacts, logs, and protocol/service knowledge as a single source of truth for testing decisions.
2. **Delta-driven change detection** on the knowledge graph that identifies meaningful changes (e.g., software updates, KPI drifts, configuration changes, topology updates) and triggers downstream testing workflows only when warranted.
3. **KG-guided generative test agent with RAG** that operates on relevant KG subgraphs, augmented by retrieval from a domain knowledge base, to create, adapt, or retire tests in a context-aware manner, integrated with test execution environments and CI/CD pipelines.
4. **MCP-based integration architecture** that decouples the AI reasoning layer from underlying data stores and execution environments, improving portability and extensibility.
5. **Prototype implementation and demonstration** on a Python-based KPI monitoring application in GitLab, showing end-to-end ingestion, change detection, test generation, execution, and merge-request creation.

The remainder of this paper is organized as follows. Section 2 formalizes the problem. Section 3 describes the proposed solution. Section 4 details the system architecture. Section 5 presents use cases. Section 6 reports implementation and experimental results. Section 7 discusses advantages and limitations. Section 8 surveys related work. Section 9 concludes.

# 2 Problem Statement

## 2.1 Formal Problem Definition

Let $S(t)$ denote the state of a telecom system at time $t$, comprising software artifacts, network topology, configuration parameters, and observable KPIs. Let $T(t)$ denote the test suite at time $t$. The objective is to maintain the property:

$$Coverage(T(t), S(t)) \geq \theta \quad \forall t$$

where $\theta$ is a minimum acceptable coverage threshold. Existing approaches compute $T(t_0)$ for a fixed snapshot $S(t_0)$ and do not update $T$ as $S$ evolves. The result is coverage degradation: $Coverage(T(t_0), S(t)) \ll \theta$ for $t \gg t_0$.

## 2.2 Qualitative Limitations of Existing Approaches

Existing automated testing approaches exhibit the following limitations:

- **Static snapshots:** Test cases are generated based on a snapshot of the system and remain static unless manually updated. In rapidly evolving telecom environments, this leads to stale tests that may no longer exercise the right code paths or network behaviors.
- **Rapid system evolution:** Underlying systems (codebases, network configurations, KPIs) change frequently. In 5G environments, network slice parameters, NF configurations, and software versions can change on timescales of hours, making static test suites obsolete quickly.
- **Lack of fine-grained change detection:** There is no systematic mechanism to detect deltas at the level of code, configuration, topology, or KPI drift and selectively update only the affected tests. Most tools operate on coarse-grained triggers (e.g., a new software release) rather than semantic changes within a release.
- **Inefficient regeneration:** Regenerating entire test suites is inefficient, time-consuming, and prone to errors, especially in large-scale telecom environments. Large operator networks may host hundreds of NFs and thousands of interdependent configuration parameters.
- **Lack of domain grounding:** LLM-based test generators that lack access to domain-specific context (protocol specifications, KPI definitions, service dependencies) produce generic tests that may be syntactically valid but semantically inadequate for telecom use.
- **Limited feedback integration:** Existing tools do not typically feed test execution results back into their knowledge representations, forgoing the opportunity to learn from failures and improve future test quality.

These limitations are particularly problematic in telecom systems, where both software and live networks evolve continuously and where test coverage must remain tightly aligned with the current system state to meet stringent SLA and compliance requirements.

# 3 Proposed Solution

We propose a context-aware, dynamically adapting test generation framework built around three core principles, extended with two architectural enhancements over the baseline concept.

### 3.1 Knowledge Graph as a Single Source of Truth

All relevant information about the system under test, including topology, software artifacts, configurations, KPIs, logs, protocols, and existing test cases, is ingested into a unified knowledge graph that is incrementally updated to reflect the latest system state. We represent the KG as a labeled property graph $G = (V, E, \lambda, \mu)$ where $V$ is the set of nodes (entities), $E \subseteq V \times V$ is the set of edges (relationships), $\lambda: V \cup E \rightarrow L$ assigns labels, and $\mu: V \cup E \rightarrow {}^{\partial}B$ assigns property maps. This formalism naturally accommodates heterogeneous telecom entities (NFs, cells, UEs, services, code artifacts) and their evolving relationships.

Ontological alignment with established telecom standards, particularly TM Forum's Information Framework (SID) [6] and 3GPP's management data models [7], ensures that the KG schema reflects industry-standard concepts rather than idiosyncratic internal representations, improving interoperability and reuse.

### 3.2 Delta-Driven Change Detection

A delta engine continuously monitors the knowledge graph for meaningful changes at different granularities (node addition/removal, edge addition/removal, property value change, or KPI drift beyond a configurable threshold). Formally, a delta $\Delta(t_1, t_2) = G(t_2) \bowtie G(t_1)$ captures the symmetric difference between successive KG states, filtered by a relevance function $\rho$ that discards low-significance changes (e.g., minor metric fluctuations within expected bounds).

Only detected, relevance-filtered deltas are propagated to the reasoning layer, ensuring that the system focuses on impacted areas and avoids unnecessary test regeneration. This selectivity is critical for scalability: in a large operator network, the majority of KG state is stable at any given moment, and wholesale regeneration would be computationally prohibitive.

### 3.3 KG-Guided Generative Test Agent with RAG

A generative AI reasoning engine analyzes detected deltas in the context of the KG. Unlike naive prompt-based test generation, our agent conditions generation on: (a) the relevant KG subgraph capturing the changed entities and their neighbors, (b) retrieved artifacts from a vector-indexed knowledge base (RAG), including protocol specifications, past test cases, KPI definitions, and known failure modes, and (c) execution history and feedback from prior test runs stored in the KG.

The agent runs on a Model Context Protocol (MCP) server, which standardizes how context is provided to and actions are taken by the LLM. MCP's tool-calling and resource-access abstractions allow the agent to query the KG, access GitLab APIs, invoke test runners, and commit results without requiring custom integration code for each action [8].

Figure 1 illustrates the high-level flow: data ingestion into the KG, delta-driven change detection, KG-guided reasoning with RAG, test generation, execution, and feedback into the KG.

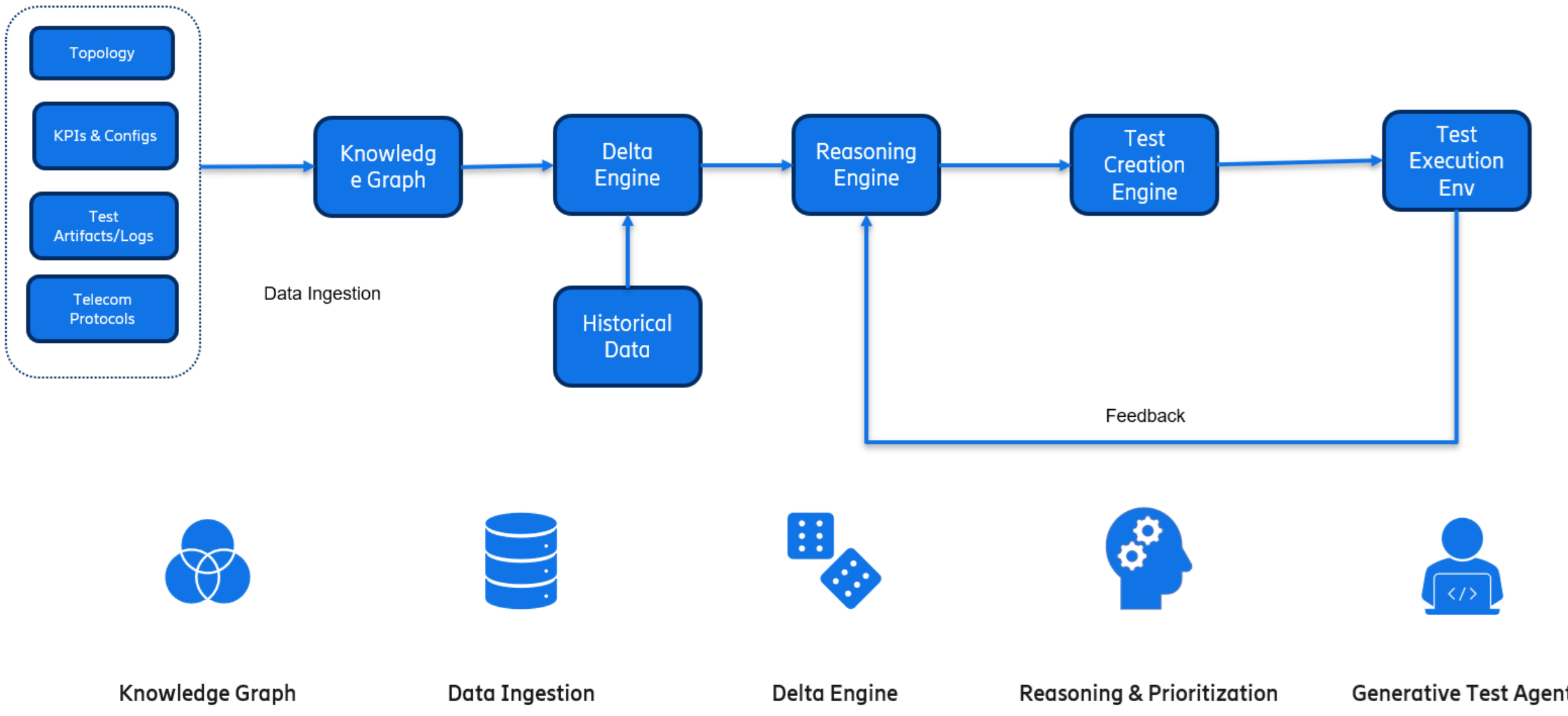

*Figure 1: High-level architecture of the context-aware generative AI test generation framework*

# 4 System Architecture and Workflow

## 4.1 Data Ingestion

Data ingestion components collect updates from multiple sources:

- **Software repositories** (e.g., GitLab/GitHub): code files, functions, classes, documentation, commit messages, and branch/merge metadata. Ingestion is event-driven, triggered by webhooks on push events or merge request state changes.
- **CI/CD pipelines:** build artifacts, test logs, coverage reports, and deployment manifests. Pipeline events are consumed via CI/CD system APIs (e.g., GitLab CI events, Jenkins webhooks).
- **Network monitoring and OSS/BSS systems:** KPIs, topology exports, configuration data from network management systems, O1/O2 interfaces in O-RAN environments, and NETCONF/YANG configuration datastores.
- **Test repositories:** existing test scripts, their parameters, execution results, and coverage metadata.
- **External knowledge sources:** 3GPP specification extracts, vendor documentation, and internal runbooks, indexed for RAG retrieval.

Each update is written to the knowledge graph with provenance metadata (source system, timestamp, confidence score), allowing the system to trace where facts originated, assess their freshness, and reconcile conflicting updates over time using a defined precedence policy. Updates from heterogeneous sources are normalized via source-specific adapters. An ingestion queue (e.g., Apache Kafka) decouples source event rates from KG write throughput, enabling replay and exactly-once semantics for critical provenance records.

## 4.2 Knowledge Graph

The knowledge graph stores the following entity and relationship types:

- **Software structures:** files, modules, functions, classes, and their dependency, inheritance, and call relationships. Abstract Syntax Tree (AST) analysis enriches code nodes with structural metadata (cyclomatic complexity, parameter signatures, return types) relevant to test generation.
- **Network entities:** NFs, cells, base stations, UEs, links, slices, and services, together with their connectivity, adjacency, and dependency relationships.
- **KPIs and metrics:** latency, availability, throughput, call drop rates, handover success rates, RSRP values, and domain-specific indicators, each linked to the entity they characterize, the threshold values defined in SLA or configuration, and historical time-series references.
- **Test cases and historical execution results:** test scripts, their input parameters, expected outcomes, actual outcomes, and execution timestamps. Each test is linked to the software features, network entities, or KPI definitions it exercises.
- **Relationships between system components and tests:** coverage edges explicitly capturing which tests cover which features, services, or KPIs, enabling gap analysis.
- **Protocol and domain knowledge nodes:** representations of 3GPP procedure descriptions, YANG model nodes, and service-level specifications, linked to relevant test cases and KPI definitions.

The KG is implemented on a graph database engine (e.g., Neo4j or Falkor DB) chosen for its support of property graph queries (Cypher or SPARQL), transactional writes, and graph traversal performance at telecom-relevant scales.

### 4.3 Delta Engine

The delta engine compares successive KG states to identify relevant changes. Its operation proceeds in three stages:

**Stage 1: Snapshot comparison.** On each ingestion cycle, or triggered by an explicit event (e.g., a branch push), the delta engine computes the difference between the newly ingested subgraph and the previously persisted KG state. Changes are classified as:

- *Structural deltas:* addition or removal of nodes (new features, retired NFs) or edges (new dependencies, removed links).
- *Property deltas:* changes to node or edge attributes (updated configuration parameters, modified function signatures).
- *KPI drift:* KPI values exceeding configurable drift thresholds, indicating potential performance degradation.
- *Coverage gaps:* new or modified entities that have no associated test cases, identified by absence of coverage edges.

Coverage gaps are identified by querying the KG for entities with no outgoing `TESTED_BY` edges. The following Cypher query illustrates gap detection in the prototype:

```
MATCH (f:Feature)
WHERE NOT (f)-[:TESTED_BY]->(:TestCase)
RETURN f.name AS FeatureName, f.impact_score AS Impact
```

```
ORDER BY f.impact_score DESC
```

The query returns uncovered features ranked by impact score, ensuring the reasoning engine prioritises test generation for high-SLA components (e.g., AMF registration flows, network slice KPIs) before lower-priority entities. Once a test is generated and passes execution, a new `TestCase` node and `[:TESTED_BY]` edge are written back to the KG, removing the feature from future gap reports and closing the coverage loop.

**Stage 2: Relevance filtering.** Not all changes warrant test updates. The relevance filter applies domain rules (e.g., "ignore cosmetic documentation changes," "flag any KPI threshold modification") and a learned significance model that scores deltas based on historical correlation with defect occurrence and test failure rates.

**Stage 3: Impact propagation.** For each significant delta, the engine traverses the KG to identify the transitive impact set, i.e. the set of entities (features, services, test cases) likely affected by the change. This graph traversal replaces manual impact analysis, which is error-prone and time-consuming in large systems. The output is a structured delta report: a ranked list of (delta, impact set, recommended action) tuples passed to the reasoning layer.

### 4.4 Reasoning and Generative Test Creation

The generative AI reasoning engine receives delta reports and operates in the following sequence:

**Context assembly.** For each delta in the report, the engine extracts the relevant KG subgraph (the changed entity, its neighbors within k hops, associated test cases, and historical execution results). This subgraph is serialized into a structured context representation suitable for LLM consumption.

**RAG augmentation.** The assembled context is augmented by retrieving relevant artifacts from the vector knowledge base: protocol specification paragraphs semantically similar to the changed entities, example test cases from historical repositories, KPI definition documents, and known failure patterns. Retrieval uses dense vector similarity search (using embeddings from a domain-fine-tuned encoder) followed by relevance re-ranking.

**Reasoning and decision.** The LLM, prompted with the assembled context and retrieved artifacts, performs three reasoning steps: (i) *impact assessment*: identifies which services, flows, or protocol procedures are affected by the delta; (ii) *action selection*: decides whether to create a new test, update an existing test, retire an obsolete test, or take no action; and (iii) *test synthesis*: generates a concrete test script or modification, grounded in the KG context and retrieved examples.

**Prompt engineering.** Prompts follow a structured template: system role definition (emphasizing telecom domain expertise and safe test generation), context section (serialized KG subgraph), retrieved knowledge section (RAG artifacts), instruction section (explicit action requested), and output format specification (test script in target language and framework, e.g., Python/pytest). Chain-of-thought prompting is used for the impact assessment step to improve reasoning quality [9].

**Output validation.** Generated test scripts are subjected to static analysis (syntax checking, import validation, assertion structure verification) before execution, filtering out malformed outputs before they reach the test runner.

### 4.5 MCP Server Integration

The generative agent operates as a client of an MCP server that exposes the following tools and resources:

- **KG query tool:** executes parameterized graph queries (Cypher/SPARQL) and returns structured results.
- **GitLab API tool:** reads file contents, creates branches, commits files, and opens merge requests.
- **Test runner tool:** executes test scripts in an isolated environment and returns results.
- **RAG retrieval tool:** queries the vector knowledge base and returns ranked document chunks.
- **KG write tool:** persists new facts (generated tests, execution results, delta records) back to the KG.

This tool-based architecture decouples the LLM from the specifics of each integration, allowing new data sources or execution environments to be added by registering additional MCP tools without modifying the core reasoning logic.

### 4.6 Test Execution and Feedback Loop

Generated or updated tests are executed in a test execution environment (containerized for reproducibility and safety). Execution results, including successes, failures, and coverage metrics, are written back into the KG as additional test artifact nodes linked to the corresponding feature and test nodes. This closes the loop and allows the system to:

- **Learn from prior failures:** test cases that consistently fail against correct implementations are flagged for review; the failure context informs future generation to avoid similar issues.
- **Refine relevance scoring:** deltas associated with test failures receive higher significance scores in future delta engine cycles, improving prioritization.
- **Maintain alignment with observed system performance:** test parameters (e.g., KPI thresholds) are updated when execution results reveal that current thresholds no longer reflect actual system behavior.
- **Track test health over time:** the KG maintains a full history of test creation, modification, retirement, and execution results, supporting audit trails and compliance reporting.

## 5 Use Cases

### 5.1 Software System Use Case: GitLab KPI Monitoring Application

In a software development scenario, we consider a Python-based KPI monitoring application maintained in GitLab. The application provides features to calculate cell availability as a percentage, check whether RSRP signal strength is within acceptable bounds, and verify that call drop rate does not exceed a configured threshold.

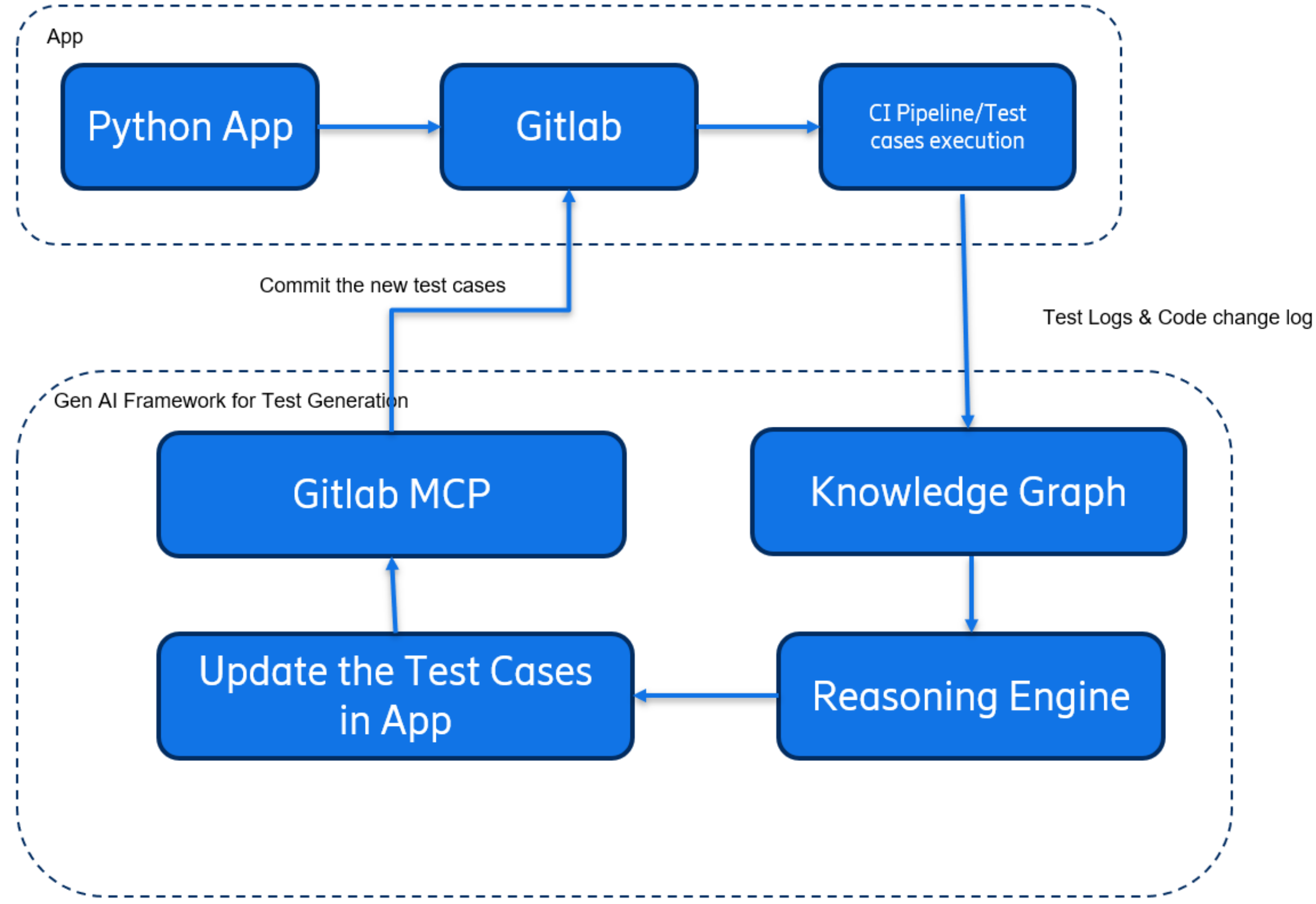

*Figure 2: Software System Test Case Generation Use Case*

**Initial ingestion and KG state.** Data from the develop branch, including code files, functions, and classes, is ingested into the graph database. Each feature is represented as a node (type: Feature), each function as a node (type: Function), and existing test cases as linked nodes (type: TestCase) with coverage edges. Initially, three feature nodes and their corresponding test cases are present in the KG.

**Feature extension and change detection.** New features are implemented and pushed into a new feature branch: (i) check handover success rate, (ii) validate that network latency meets SLA requirements, and (iii) calculate network throughput in Mbps, along with partial test cases for these features. The delta engine detects these changes by comparing the feature branch state with the current KG state and identifies three new Feature nodes, three new Function nodes with updated parameter signatures, two coverage gaps (new features with no associated complete TestCase nodes), and one partial test case requiring extension.

**Generative reasoning and test creation.** The reasoning engine is invoked: it initializes the GitLab and KG MCPs, clones the develop branch into the agent workspace, identifies the feature branch containing the new features, reads the software specification from README.md, and queries the KG for current test coverage. Based on the detected deltas and existing tests, it generates missing or updated test cases, runs them locally in the test execution environment, and evaluates their outcomes.

**Branching and human review.** After successful local execution, the agent creates a new branch dedicated to auto-generated tests, commits the new test cases with structured commit messages referencing the delta report, and raises a merge request from this branch to the feature branch for human approval. Once approved and merged into the feature branch and subsequently into develop, the KG is updated, now showing six feature nodes (three original plus three new) with associated test cases. This demonstrates continuous, context-aware test maintenance with minimal manual effort.

### 5.2 Telecom Network Use Case

For live telecom networks, the system ingests network topology, KPIs, and configuration data into the knowledge graph via O1/O2 interface adapters (for O-RAN environments) and NETCONF/RESTCONF adapters (for traditional NF management). When changes such as KPI anomalies, performance degradation, or configuration updates occur, the delta engine flags the affected subgraphs, and the generative agent reacts by generating or adapting tests relevant to impacted services or flows (e.g., call setup, throughput, latency), while avoiding unnecessary tests on unaffected parts of the network.

The same framework applies to a range of additional network scenarios:

- *KPI anomaly-driven testing:* a drop in handover success rate below the SLA threshold triggers generation of targeted handover procedure tests, focused on the affected cell pairs and their configuration.
- *Network slice instantiation:* creation of a new network slice triggers generation of slice-specific SLA validation tests, including latency, throughput, and isolation tests, parameterized with the new slice's configuration from the KG.
- *NF software update:* a new version of the AMF (Access and Mobility Management Function) is deployed; the delta engine detects the version change and triggers regeneration of registration and mobility procedure tests for the updated NF.
- *Microservices and Kubernetes environments:* for cloud-native NFs, the agent produces API-level tests (REST/gRPC endpoint validation), Kubernetes readiness/liveness probe checks, and resource consumption boundary tests.
- *IoT device management:* for large-scale IoT deployments, the agent generates device probe tests and protocol conformance checks suited to constrained device capabilities.

## 6 Implementation and Results

### 6.1 Implementation Details

The prototype was implemented on top of a Python 3.x-based KPI monitoring application hosted in GitLab, which exposes three initial features: cell availability calculation, RSRP signal-strength validation, and call-drop-rate checks. The code and configuration from the `develop` branch are ingested into a graph database (Falkor DB or Neo4j) to construct the knowledge graph, with nodes representing files, functions, features, and test cases, and edges capturing relationships such as "implements", "tested-by", and "depends-on". A GitLab MCP integration allows the reasoning engine to clone branches, read specifications (e.g., `README.md`), and commit updated tests back to the repository. The generative AI reasoning engine (LLM-based) runs behind an MCP server, receives detected deltas from the KG, retrieves the relevant subgraph, and synthesizes or adapts test scripts in Python's standard testing style. A simple scheduler (cron-style) is used to keep the KG synchronized with the repository and CI/test results, although in the current prototype the ingestion and reasoning pipeline is triggered on demand rather than fully automated on every commit.

### 6.2 Prototype Setup

We implemented a proof-of-concept prototype of the proposed framework on a Python-based KPI monitoring application hosted in GitLab.

The application initially exposes three features: (i) calculate cell availability as a percentage, (ii) verify that RSRP signal strength is within an acceptable range, and (iii) check whether the call drop rate exceeds a configured threshold.

The prototype includes a graph database (Falkor DB or Neo4j) to store the knowledge graph, a GitLab MCP integration to interact with the repository, a generative AI reasoning engine, and a local test execution environment.

In the initial state, data from the `develop` branch—code files, functions, and classes—is ingested into the knowledge graph so that each feature appears as a node with associated test-case nodes, resulting in three feature nodes and their corresponding tests.

### 6.3 Demonstration Procedure

To emulate typical evolution of a telecom KPI application, we added three new features in a dedicated feature branch: (i) check handover success rate, (ii) validate that network latency meets SLA requirements, and (iii) calculate network throughput in Mbps.

Partial test cases for these features were also committed to the feature branch to reflect incomplete manual test maintenance.

The following sequence was executed:

1. The agent was started and initialized the GitLab and Falkor DB MCPs.
2. It created a plan, cloned the `develop` branch into its workspace, and discovered the feature branch containing the new functionality.
3. It read the software specification from the `README.md` file and queried the knowledge graph to obtain the current state of test cases for the three existing features.
4. It compared the feature branch contents with the KG state, identified gaps in test coverage for the new features, and generated the missing test cases.
5. The newly generated tests were executed locally in the test execution environment.
6. After validation, the agent created a new branch, committed the generated tests, and raised a merge request from this `auto-generated-test` branch to the feature branch.
7. Finally, after the engineer approved and merged the request into the feature branch and then into `develop`, the graph database was updated and showed six feature nodes (the three original plus three new), each with associated test cases.

This end-to-end sequence demonstrates the closed loop from code changes to knowledge-graph updates, test generation, execution, and integration into the main branch.

### 6.4 Observed Results

The demonstration highlights several concrete outcomes:

- **Automatic coverage of new features:** After the feature branch was created with three additional KPI-related functions, the agent correctly identified that only the new features lacked complete tests and generated the missing cases, without modifying tests for unchanged functionality.
- **Reduction of manual test authoring:** The engineer's role was limited to reviewing and approving the merge request containing the generated tests; there was no need to hand-write test cases for the new features.
- **Consistent KG–repository state:** Before changes, the KG contained three feature nodes with tests; after the demo workflow, the KG contained six feature nodes with associated test cases, matching the updated develop branch.
- **End-to-end autonomy:** Once triggered, the agent autonomously ingested data, planned its actions, generated tests, executed them, and produced a merge request, demonstrating how the framework can be integrated into existing GitLab-centric development flows.

Although the demonstration focuses on a single Python application, it shows that the proposed architecture can handle real repository structures and CI-friendly workflows, rather than only synthetic examples.

### 6.5 Discussion and Limitations

These results are based on a demonstration-scale prototype rather than a large-scale empirical study. We did not yet measure quantitative metrics such as time saved per feature, coverage improvement, or fault-detection rates; instead, the focus was on validating the functional loop from change detection to test generation and integration.

Future experimental work will include controlled studies comparing manual versus KG-guided generative testing on larger telecom software and network scenarios, and measuring: (i) reduction in human effort for test maintenance, (ii) impact on defect detection, and (iii) the overhead of maintaining and querying the knowledge graph at scale.

## 7 Advantages and Considerations

Compared to traditional automated testing approaches, the proposed framework offers several advantages:

1. **Continuous alignment between tests and the live system.** The knowledge graph is incrementally updated, and the delta engine monitors changes, so tests are created or adapted only when relevant facts change. This keeps test suites aligned with the real software and network state without requiring periodic manual curation.
2. **Reduced manual effort through selective, delta-driven updates.** Instead of regenerating entire test suites, the system selectively updates only the tests affected by detected deltas, significantly reducing manual effort and review overhead.
3. **More accurate and safer tests grounded in real context.** Generated tests are grounded in real system relationships, protocol models, KPIs, and historical failures captured in the KG, improving their relevance and reducing the risk of generating tests that exercise unreachable code paths or impossible network states.

4. **Highly targeted, context-aware test scripts.** Conditioning the generative agent on precise KG subgraphs and RAG-retrieved domain knowledge leads to highly targeted tests focused on impacted services and flows, avoiding generic or redundant test cases.
5. **Faster test cycles and improved development velocity.** By automating test maintenance and tightly coupling it with CI/CD workflows, the framework accelerates test cycles, reduces the feedback latency between code changes and test results, and supports faster deployment of new features—a critical advantage in competitive 5G service deployment contexts.
6. **Persistent learning and continuous improvement.** The feedback loop from test execution into the KG creates a continuously improving test intelligence base. Over time, the system accumulates knowledge of which test patterns are most effective for which change types.

At the same time, important considerations remain:

- **Quality control:** human review (e.g., via merge requests) is still essential, particularly in safety- or SLA-critical telecom environments. The framework is designed to support, not replace, human judgment on test quality and safety.
- **KG completeness and correctness:** the effectiveness of the approach depends on the coverage and accuracy of the KG; missing or stale data may result in incomplete or misdirected tests. KG health monitoring and staleness alerting are necessary operational components.
- **LLM reliability and hallucination:** LLMs can generate syntactically valid but semantically incorrect tests. The static analysis validation layer and mandatory human review mitigate this risk, but fine-tuning on telecom-domain test corpora may be needed for production deployment.
- **Scalability:** scaling KG operations and delta detection to very large telecom networks requires careful engineering—including KG partitioning, incremental indexing, and distributed delta computation—beyond the scope of the current prototype.
- **Security and access control:** the MCP agent has broad access to source repositories, test environments, and potentially production network APIs. Robust authentication, authorization, and audit logging are essential.
- **Prompt injection and adversarial inputs:** care must be taken to sanitize KG-derived content before inclusion in LLM prompts, as maliciously crafted repository content or network configuration values could influence LLM behavior in unintended ways.

## 8 Related Work

Prior work in AI-powered test automation, telecom testing, knowledge-graph-based network management, and generative AI for software engineering provides valuable foundations but does not address dynamic, change-driven test generation in the unified manner proposed here.

### 8.1 AI-Powered Test Automation

Commercial tools such as Testim (Tricentis) [10] and Mabl use machine learning to stabilize UI test selectors and reduce maintenance for web applications, but they do not incorporate knowledge graphs, telecom domain knowledge, or continuous change detection for network-level testing. Academic work on LLM-based unit test generation [2, 3]—including ChatUniTest, TestPilot, and GitHub Copilot-assisted testing—demonstrates the feasibility of LLM test synthesis for software, but these approaches are code-only, single-shot, and do not adapt to system evolution.

### 8.2 Telecom Testing Platforms

Platforms such as Spirent's Landslide and Keysight's IxLoad support rich network testing for 4G/5G protocols, offering scripted test case libraries and performance benchmarking capabilities [11]. However, these platforms operate on manually authored test libraries and lack automated detection of network changes to drive dynamic test regeneration. They provide the execution substrate but not the intelligent test synthesis layer proposed here.

### 8.3 Knowledge-Graph-Based Network Management

Research systems such as SeaNet [4] and ITSM-KG/NORIA [5] demonstrate how knowledge graphs can support autonomic network management, incident handling, root cause analysis, and risk assessment. These systems establish the value of KG-based representations for network operations but do not focus on automatic test case generation. The ETSI ZSM (Zero-touch network and Service Management) framework [12] similarly advocates for knowledge-driven autonomous management but does not address test generation specifically.

### 8.4 KG-Based Test Generation in Software Engineering

Nayak et al. [13] show how KGs constructed from software requirements and domain documents can be used to derive test cases, demonstrating improved coverage of requirement-level behaviors compared to manual test design. However, this approach is static (applied once to a fixed requirements document), does not address continuous change detection, and has not been applied to telecom-specific contexts.

### 8.5 Security-Focused KG Approaches

Research on KG-based penetration testing and network security [14] uses KGs to model attack surfaces and generate attack scenarios for vulnerability assessment. While this demonstrates KG-guided test generation in a network context, it is limited to security domains, does not cover general telecom service correctness testing, and does not incorporate generative AI for test synthesis.

### 8.6 Retrieval-Augmented Generation for Software Tasks

RAG has been applied to code generation [15], documentation synthesis, and bug localization tasks, demonstrating that grounding LLM outputs in retrieved domain-specific artifacts improves relevance and reduces hallucination rates compared to pure parametric generation. Our framework extends this principle to test generation, using RAG to ground generated tests in telecom protocol specifications and historical test artifacts.

### 8.7 Model Context Protocol

The Model Context Protocol (MCP) [8], introduced by Anthropic, provides a standardized interface for LLM agents to interact with external tools and data sources. While MCP has been used in general agentic AI applications, its application to telecom test automation and KG-guided reasoning is novel to this work.

### 8.8 Patent Literature

Patent literature on AI-based testing and knowledge-graph construction for cloud and network systems (e.g., US10515002B2, CN110347603B, EP4482193A4, US20220245470A1) addresses individual components—KG construction, AI-based test selection, or network monitoring—but consistently lacks either dynamic KG updates, generative test synthesis, or telecom-specific adaptation in a unified framework. Our contribution is precisely this unification.

### 8.9 Summary of Gaps

Table 1 summarizes the coverage of related approaches across the key dimensions of our framework.

| **Approach** | **Continuous KG** | **Delta Detection** | **Generative Tests** | **Telecom Domain** | **RAG Aug.** |
|---|---|---|---|---|---|
| Testim / Mabl | ✗ | ✗ | ✗ | ✗ | ✗ |
| LLM Test Gen (ChatUniTest etc.) | ✗ | ✗ | ✓ | ✗ | Partial |
| Spirent / Keysight | ✗ | ✗ | ✗ | ✓ | ✗ |
| SeaNet / NORIA | ✓ | Partial | ✗ | ✓ | ✗ |
| Nayak et al. KG-Test | Partial | ✗ | ✗ | ✗ | ✗ |
| Security KG Testing | Partial | ✗ | Partial | Partial | ✗ |
| **This work** | ✓ | ✓ | ✓ | ✓ | ✓ |

*Table 1: Comparison of related approaches across key framework dimensions.*

Our approach bridges these gaps by combining continuous change detection, knowledge graph integration, RAG-augmented reasoning, and generative adaptation of tests into a single, unified framework specifically designed for telecom environments.

## 9 Conclusion and Future Work

This paper presented a context-aware generative AI framework for automated telecom test script generation that adapts continuously to system changes. By leveraging a knowledge graph as a living representation of the system, delta-driven change detection, RAG-augmented reasoning, and a generative AI agent operating via the Model Context Protocol, the proposed approach keeps test suites relevant, focused, and efficient for both software systems and live telecom networks.

Our prototype on a Python-based KPI monitoring application illustrates the feasibility of end-to-end integration, from KG ingestion and change detection to test generation, execution, and merge-request creation, within an existing GitLab-based development workflow. Preliminary results validate the functional loop from change detection to test generation and integration with minimal manual intervention.

Some areas for future work include the following:

- **Deeper semantic reasoning over telecom protocols:** incorporating formal protocol state machine representations (e.g., 3GPP NAS procedure state machines) into the KG to enable generation of protocol-conformance tests that exercise specific state transitions.
- **Tighter integration with live network controllers:** connecting the ingestion pipeline to real-time network telemetry streams (e.g., via O-RAN O1 streaming telemetry) for sub-minute delta detection and response in production environments.
- **Empirical evaluation at scale:** rigorous measurement of fault detection improvement, test-suite efficiency (reduction in redundant tests), and test cycle acceleration in a large-scale operator environment, including comparison with baseline approaches.
- **Multi-modal KG enrichment:** incorporating unstructured knowledge sources (vendor documentation, incident tickets, field engineer notes) into the RAG knowledge base via automated extraction pipelines.
- **Federated KG architectures:** exploring distributed KG deployments that respect data sovereignty requirements across multi-operator or multi-vendor environments, enabling cross-domain test coordination without centralizing sensitive network data.
- **LLM fine-tuning on telecom test corpora:** systematically fine-tuning the generative model on large collections of telecom-domain test scripts and specifications to improve domain-specific generation quality and reduce hallucination rates.
- **Adversarial robustness:** evaluating and hardening the framework against prompt injection and adversarial inputs that could influence test generation in unintended ways.

By tightly coupling knowledge graphs, delta-driven change detection, RAG-augmented reasoning, and generative AI, we take a step toward intelligent, self-adapting test systems for modern telecom and cloud-native software environments, which are systems that can keep pace with the relentless evolution of the networks they validate.